%% file: parallel_MAP.tex
\documentclass[twocolumn]{autart}     
\usepackage{graphicx}          
\usepackage{amsmath}   
\usepackage{amsfonts}
\usepackage{pgfplots}
\pgfplotsset{compat=1.18}  
\usepackage{cite}
\usepackage[utf8]{inputenc}

\usepackage{xcolor}

\usepackage{xpatch}
\makeatletter 
\xpatchcmd{\runningauthor@fmt}{\global\edef}{\protected@xdef}{}{}
\xpatchcmd{\runningauthor@fmt}{\global\edef}{\protected@xdef}{}{}
\xpatchcmd{\author@fmt}{\edef}{\protected@edef}{}{}
\def\@xnamedef#1{\expandafter\protected@xdef\csname #1\endcsname}
\def\ead@au#1{\protected@edef\@ead@au{#1}}
\def\add@xtok#1#2{\begingroup
  \protected@xdef\@act{\global\noexpand#1{\the#1#2}}\@act
\endgroup}
\def\no@harm{}
\makeatother

\begin{document}

\begin{frontmatter}

\title{Temporal parallelisation of continuous-time maximum-a-posteriori trajectory estimation\thanksref{footnoteinfo}}

\thanks[footnoteinfo]{Corresponding author H. Razavi.}

\author[Aalto]{Hassan Razavi}\ead{hassan.razavi@aalto.fi},    
\author[UPM]{\'Angel F. Garc\'ia-Fern\'andez}\ead{angel.garcia.fernandez@upm.es
},    
\author[Aalto,ELLIS]{Simo S\"arkk\"a}
\ead{simo.sarkka@aalto.fi}  

\address[Aalto]{Department of Electrical Engineering and Automation, Aalto University, Espoo, Finland}               

\address[UPM]{Information Processing and Telecommunications Center, Universidad Polit\'ecnica de Madrid, Madrid, Spain}             
\address[ELLIS]{ELLIS Institute Finland, Aalto University, Espoo, Finland}
          

\begin{abstract}                         
This paper proposes a parallel-in-time method for computing continuous-time maximum-a-posteriori (MAP) trajectory estimates of the states of partially observed stochastic differential equations (SDEs), with the goal of improving computational speed on parallel architectures. The MAP estimation problem is reformulated as a continuous-time optimal control problem based on the Onsager--Machlup functional. This reformulation enables the use of a previously proposed parallel-in-time solution for optimal control problems, which we adapt to the current problem. The structure of the resulting optimal control problem admits a parallel solution based on parallel associative scan algorithms. In the linear Gaussian special case, it yields a parallel Kalman–Bucy filter and a parallel continuous-time Rauch--Tung--Striebel smoother. These linear computational methods are further extended to nonlinear continuous-time state-space models through Taylor expansions. We also present the corresponding parallel two-filter smoother. The graphics processing unit (GPU) experiments on linear and nonlinear models demonstrate that the proposed framework achieves a significant speedup in computations while maintaining the accuracy of sequential algorithms.
\end{abstract}

\end{frontmatter}

\section{Introduction} \label{sec:intro}

Stochastic differential equations (SDEs) commonly arise as models of dynamic phenomena in various applications such as target tracking, communications, and finance \cite{Jazwinski:1970, Van-Trees:1968, Van-Trees:1971, sarkka2019applied, Sarkka+Svensson:2023}. In many of these applications, the state of the dynamic system is not observed directly, but instead, we obtain noisy measurements of the underlying phenomenon. In such partially observed systems, an important task is to estimate the underlying state from noisy measurements. One way to estimate the state is to compute the maximum-a-posteriori (MAP) estimate of the state trajectory, which roughly corresponds to determining the most probable state trajectory given the noisy measurements \cite{Dutra:2014, dutra2017joint, sarkka2019applied}. 

In the context of discrete-space systems, the classical algorithm for computing MAP trajectory estimates is the Viterbi algorithm \cite{Viterbi:1967, Larson+Peschon:1966, Sarkka+Svensson:2023}, which can be interpreted as a dynamic programming method \cite{Bellman:1957} for computing the most likely state sequence. The Viterbi algorithm can also be derived as a solution to a suitably defined optimal control problem \cite{Omura:1969, sarkka2023temporal} which is solved using dynamic programming. The classical Viterbi algorithm proceeds in two stages: a forward pass, which propagates the probabilities of the states using the optimal probabilities up to the previous time step, and a backward pass, which reconstructs the most likely sequence of states by tracing back through the states that maximize the probability at each step \cite{Larson+Peschon:1966, Viterbi:1967}. Another perspective views the Viterbi algorithm as a max-product procedure \cite{wymeersch2007iterative}, in which independent forward and backward passes are performed and then combined to determine the most likely sequence of states. However, both algorithm types are sequential, that is, not parallel in time.

Recently, parallel-in-time algorithms to compute MAP estimates in the discrete-space setting (i.e., parallel Viterbi algorithms) have been proposed in \cite{hassan2021temporal, sarkka2023temporal}. The paper \cite{hassan2021temporal} is particularly concerned with parallelising the max-product formulation of the Viterbi algorithm using parallel scan algorithms \cite{Blelloch:1989}. Also, based on parallel associative scans, the paper \cite{sarkka2023temporal} considered the parallelisation of the optimal control formulation of the Viterbi algorithm by leveraging the algorithms developed in \cite{Sarkka:2023}. These parallel algorithms have so far only been formulated for discrete-time dynamic systems.

In the discrete-time, continuous-space setting for linear-Gaussian models, the MAP trajectory estimate can be obtained using the RTS smoother. Furthermore, although the original Viterbi algorithm \cite{Viterbi:1967, Larson+Peschon:1966} was developed for discrete-valued state spaces, as an algorithm, it also equally applies to both discrete and continuous state spaces \cite{Sarkka+Svensson:2023}, and in the linear Gaussian case, its classical version reduces to the Rauch--Tung--Striebel (RTS) smoother and the max-product version reduces to the two-filter smoother. For nonlinear cases, one approach to approximate the MAP trajectory is to use the iterated extended Kalman smoother (IEKS), which corresponds to Gauss–Newton optimisation \cite{bell1994iterated}, and also sigma point Kalman smoother\cite{Sarkka+Svensson:2023}. A parallel-in-time version of the RTS smoother was proposed in \cite{Sarkka:2021}, and a parallel version of the IEKS and sigma-point Kalman smoother was presented in \cite{yaghoobi2021parallel}. 

In continuous-time stochastic system with continuous spaces, the MAP estimation problem corresponds to the minimisation of the Onsager--Machlup functional \cite{Dutra:2014, dutra2017joint, zeitouni1987maximum}. This can be seen as a continuous-time analogue of the discrete-time Viterbi algorithm. As discussed in \cite{Dutra:2014}, the classical references \cite{bryson1963, Jazwinski:1970} often solve the minimum energy problem, instead of the MAP estimation problem defined by the Onsager--Machlup functional. However, the difference is small, a single divergence term. For linear-affine systems, the MAP estimate is already computed by the continuous RTS smoother, whose its forward pass is the classical Kalman-Bucy filter \cite{sarkka2019applied}. For non-linear systems, numerical methods have been developed by using linearisation methods \cite{Jazwinski:1970, sarkka2019applied}. However, all these methods are sequential in time, and this paper aims to develop their parallel-in-time counterpart. For this purpose, we will leverage the parallelisation of the Hamilton--Jacobi--Bellman (HJB) equation and continuous-time linear quadratic regulator (LQT) developed in \cite{sarkka2024temporal}. 

The contribution of this paper is to develop a novel method for computing continuous-time MAP estimates of partially observed SDEs using parallelisation across the time domain. We first convert the Onsager--Machlup functional minimisation problem into an equivalent continuous-time optimal control problem, as was done in the discrete-time setting in  \cite{Omura:1969, sarkka2023temporal}. The general solution to the control problem is then derived based on the parallel HJB formulation \cite{sarkka2024temporal}, which gives the parallel MAP estimation algorithm. For the linear-affine case, we specialise the parallel continuous-time LQT solution \cite{sarkka2024temporal} to the MAP estimation problem, yielding a continuous-time parallel Kalman-Bucy filter and a continuous RTS smoother. We also derive the corresponding forward-backward value-function combination-based method, which yields a parallel continuous-time two-filter smoother. Taylor-series-based methods for nonlinear state-space models are also presented. Finally, we test the practical performance of the algorithms on a graphics processing unit (GPU).

\section{Continuous-time MAP estimation as an optimal control problem} \label{sec:control-map}

In this section, we first define the continuous-time stochastic state-space model, and then describe how to obtain the MAP trajectory estimate by minimising the Onsager--Machlup cost functional. Next, we reformulate the problem as a continuous-time optimal control problem, and we show how the HJB equation gives the solution to the MAP estimation problem. After that, we specialise the solutions to the linear-affine case.

\subsection{Continuous-time MAP estimate and the Onsager--Machlup functional}\label{MAP-Machlup}

The MAP estimate for a partially observed continuous-time stochastic system is the state trajectory $x^*(t)$ that maximises the posterior probability given the observed measurements (see, e.g., \cite{Dutra:2014} for a more formal definition). Let us consider a stochastic model with a hidden state $x(t) \in \mathbb{R}^{n_x}$, and an observation vector $y(t) \in \mathbb{R}^{n_y}$. The state dynamics and measurements are described by the following SDEs \cite{Jazwinski:1970, sarkka2019applied}:
\begin{equation}\label{statespace}
\begin{aligned}
    \dot{x}(t) & = f(x(t),t)  + L(t) w(t), \\
    y(t) & = h(x(t),t)  + \nu(t), 
\end{aligned}
\end{equation}
where $w(t)\in \mathbb{R}^{n_x}$ and $\nu(t)\in \mathbb{R}^{n_y}$ are white noise processes with spectral density matrices $W(t)$ and $R(t)$, respectively. The state at the initial time $t_0$ is assumed to follow an initial distribution $p(x(t_0))$. 

 One way to obtain the MAP (trajectory) estimate in continuous time is by minimising the Onsager--Machlup functional \cite{Dutra:2014, dutra2017joint, zeitouni1987maximum, sarkka2019applied}. For the time interval $[t_0,t_f]$ and a trajectory $x(t)$ defined in this time interval (also referred to as $x$), then the Onsager--Machlup cost functional is 
\begin{equation}\label{OM}
    \begin{aligned}
        &M[x] = l(x(t_{0}))+\int_{t_0}^{t_f}l_t(x,\dot{x})dt\\
        &\quad =-\log p(x(t_{0}))+ \frac{1}{2} \int_{t_0}^{t_f} \nabla \cdot f(x,t)dt \\
        &\quad + \frac{1}{2} \int_{t_0}^{t_f} \left(\dot{x}(t) - f(x,t)\right)^{\top} Q(t)^{-1}\left(\dot{x}(t) - f(x,t)\right)dt\\ &\quad  + \frac{1}{2}\int_{t_0}^{t_f} \left(y(t)-h(x,t)\right)^{\top}R(t)^{-1}\left(y(t)-h(x,t)\right) dt,
    \end{aligned}
\end{equation}
where $\nabla \cdot f(x,t)$ denotes the divergence of $f(x,t)$, and $Q(t)=L(t)W(t)L(t)^\top$ is assumed to be invertible. Note that this requirement can be relaxed by using generalizations of the Onsager--Machlup functional for singular diffusion matrices or by using pseudo-inverses \cite{aihara2002mortensen, liu2024onsager}. Furthermore, functions $f$ and $h$ need to be Lipschitz continuous in $x$; moreover, $f$ is twice differentiable in $x$, and differentiable in $t$. These assumptions ensure that the state-space in \eqref{statespace} and the cost functional in \eqref{OM} are well-defined.

In this paper, the aim is to estimate the hidden state trajectory $x(t)$ in \eqref{statespace} using the MAP trajectory estimate $x^{*}(t)$, which is given by minimising the Onsager--Machlup cost functional in \eqref{OM}.

\subsection{Optimal control problem corresponding to the Onsager--Machlup functional}\label{optimalcontrolproblem}

In this section, we reformulate the MAP trajectory estimation problem as a continuous-time optimal control problem with the same minimum as \eqref{OM}. A similar reformulation was used in \cite{sarkka2023temporal} for the Viterbi algorithm for discrete-time state-space models.

To formulate a continuous-time optimal control problem corresponding to \eqref{OM}, we need to reverse the time by defining $\tau = t_f - t$, since the initial state in the MAP estimation problem corresponds to the terminal cost in the optimal control problem. For that purpose, let us define the time-reverted quantities $\phi(\tau) = x(t_f - \tau)$, $\tilde{f}(\phi,\tau) = -f(\phi, t_f - \tau)$, $\tilde{Q}(\tau) = Q(t_f - \tau)$, $\tilde{y}(\tau) = y(t_f - \tau)$, $\tilde{h}(\phi,\tau) = h(\phi, t_f - \tau)$, and $\tilde{R}(\tau) = R(t_f - \tau)$. 

We then introduce a control signal $u(\tau) = \dot{\phi}(\tau) - \tilde{f}(\phi(\tau),\tau)$ which converts the cost functional to
    \begin{equation}\label{reversecost}
     \begin{aligned}
        M[\phi,u] 
        & = -\log(p(\phi(t_f))) -\frac{1}{2}\int_{0}^{t_f-t_0} \nabla \cdot \tilde{f}(\phi,\tau)d\tau \\ &+\frac{1}{2}\int_{0}^{t_f-t_0}  \left(\tilde{y}(\tau)-\tilde{h}(\phi,\tau)\right)^{\top}\tilde{R}(\tau)^{-1} 
        \\ & \qquad \qquad \qquad  \qquad \qquad  \times \left(\tilde{y}(\tau)-\tilde{h}(\phi,\tau)\right) d\tau\\
        &+ \frac{1}{2} \int_{0}^{t_f-t_0}u(\tau) \tilde{Q}(\tau)^{-1}u(\tau) d\tau, \\
        &:= l(\phi(t_f))+\int_{0}^{t_f-t_0}l_{\tau}(\phi,u)d\tau,
       \end{aligned}
\end{equation}
where
\begin{align}
    &\dot{\phi}(\tau)=\tilde{f}(\phi(\tau),\tau) + u(\tau), \label{controldynamic}\\
    &l(\phi(t_f))=-\log(p(\phi(t_f))) \label{parameters_oc1}\\
        &l_{\tau}(\phi,u)=\frac{1}{2}\, u^{\top}\tilde{Q}(\tau)^{-1} u
               - \frac{1}{2}\, \nabla \cdot \tilde{f}\big(\phi, \tau \big), \nonumber\\
      &+ \frac{1}{2} \left(\tilde{y}(\tau) - \tilde{h}\left(\phi, \tau\right)\right)^{\top}\tilde{R}(\tau)^{-1}\left(\tilde{y}(\tau) - \tilde{h}\left(\phi, \tau\right)\right). \label{parameters_oc2}
\end{align}
Equations \eqref{reversecost}-\eqref{parameters_oc2} define a continuous optimal control problem \cite{sarkka2024temporal}, where $\phi$ and $u$ are the state vector and the control input in continuous time and $\tau \in [\tau_0,\tau_f]$ where $\tau_0=0$ and $\tau_f=t_f-t_0$. Hence, we can find the MAP trajectory by first finding the optimal trajectory $\phi^{*}$ for the optimal control problem and then reversing the time yielding 
\begin{equation}
x^{*}(t) = \phi^{*}(t_f - t).
\label{eq:phi_reverse}
\end{equation}

\subsection{Hamilton--Jacobian--Bellman solution to the continuous MAP estimate }\label{HJB-MAP}
The HJB equation is a partial differential equation (PDE) \cite{sarkka2024temporal, kirk2004optimal}, for the value function of an optimal control problem. The value function $V(\phi,\tau)$ can be obtained as the solution to the following HJB equation: 
\begin{equation}\label{principle}
    \begin{aligned}
     -\partial_{\tau} V(\phi,\tau)&=\min_{u}\left\{l_{\tau}(\phi,u)+\partial_{\phi} V(\phi,\tau)^{\top}(\tilde{f}+u)\right\},
    \end{aligned}
\end{equation}
with the terminal cost $V(\phi,t_f)=l(\phi(t_f))$. In our case, the minimisation has a closed form solution, which gives the control law
\begin{equation}\label{optimalcontrollaw}
    u^{*}(\phi,\tau) = -\tilde{Q}(\tau)\partial_{\phi} V(\phi,\tau).
\end{equation}
The HJB equation then reduces to
\begin{equation}\label{HJB}
    \begin{aligned}
    -\partial_{\tau}& V(\phi,\tau)
    = -\frac{1}{2}\,\partial_{\phi} V(\phi,\tau)^{\top}\tilde{Q}(\tau)\partial_{\phi} V(\phi,\tau) \\
    &+ \frac{1}{2}\,\left(\tilde{y}(\tau)-\tilde{h}(\phi,\tau)\right)^{\top}\tilde{R}(\tau)^{-1}
    \left(\tilde{y}(\tau)-\tilde{h}(\phi,\tau)\right)\\
   &+ \partial_{\phi} V(\phi,\tau)^{\top} \tilde{f}(\phi,\tau) - \frac{1}{2}\,\nabla\ \cdot \tilde{f}(\phi,\tau).
    \end{aligned}
\end{equation}
Once the value function has been solved backwards in time, the optimal state trajectory can be obtained by substituting the control law \eqref{optimalcontrollaw} into \eqref{controldynamic}:
\begin{equation}\label{controllaw}
    \dot{\phi}^*(\tau)=\tilde{f}(\phi^*,\tau)+u^{*}(\phi^*,\tau),
\end{equation}
where the initial condition is $\phi^{*}(\tau_0) = \arg\min_{\phi}V(\phi,\tau_0)$. Finally, because the time variable was reversed via $\phi(\tau) = x(t_f - \tau)$, the MAP estimate in the original time variable is obtained with \eqref{eq:phi_reverse}.

\subsection{Linear quadratic tracking for the obtained optimal control problem}\label{LQT}

Let us now consider a linear-affine continuous-time state-space model, where the functions $f(\cdot)$ and $h(\cdot)$ in \eqref{statespace} are linear-affine:
\begin{equation}\label{linear-statespace}
\begin{aligned}
    \dot{x}(t) & = F(t)x(t) + c(t) + L(t) w(t), \\
    y(t) & = H(t) x(t) + r(t) + \nu(t), 
\end{aligned}
\end{equation}
and the initial distribution is Gaussian $p(x(t_0)) = \mathcal{N}(x(t_0);m_0,P_0)$. In this case, the optimal control problem defined by \eqref{reversecost}-\eqref{parameters_oc2} reduces to an LQT problem (see, e.g., \cite{sarkka2024temporal}). In particular, the corresponding LQT problem has the form
\begin{equation}\label{affine}
\begin{aligned}
    \tilde{f}(\phi,\tau)&=\tilde{F}(\tau)\phi+\tilde{c}(\tau),\\
    h(\phi,\tau)& = \tilde{H}(\tau)\phi + \tilde{r}(\tau),
    \end{aligned}
\end{equation}
where $\tilde{F}(\tau) = -F(t_f - \tau)$, $\tilde{c}(\tau) = -c(t_f - \tau)$, $\tilde{H}(\tau) = H(t_f - \tau)$, $\tilde{r}(\tau) = r(t_f - \tau)$, and the control cost matrix is $\tilde{Q}(\tau)^{-1}$ and the state cost matrix is $\tilde{R}(\tau)^{-1}$.

For the linear-affine problem, the HJB equation \eqref{principle} admits a quadratic value function \cite{sarkka2024temporal}
\begin{equation}
  V(\phi,\tau) = \frac{1}{2} \phi^\top S(\tau) \phi - v(\tau)^\top \phi
  + \text{constant},
\end{equation}
characterized by a symmetric matrix $S(\tau)$ and a vector $v(\tau)$, where the constant (though time-dependent) term independent of $\phi$ can be ignored. The functions $S(\tau)$ and $v(\tau)$ are obtained by integrating the following differential equations backwards:
\begin{equation}
    \begin{aligned}\label{2df}
        \dot{S}(\tau)& = S(\tau) \tilde{Q}(\tau)S(\tau)\\
        &-S(\tau)\tilde{F}(\tau)-\tilde{F}^{\top}(\tau)S(\tau)-\tilde{H}(\tau)^{\top}\tilde{R}^{-1}(\tau)\tilde{H}(\tau)\\
         \dot{v}(\tau)& = S(\tau)\tilde{Q}(\tau)v(\tau)-\tilde{F}(\tau)^{\top}v(\tau) \\&  +S(\tau) \tilde{c}(\tau)-\tilde{H}(\tau)^{\top}\tilde{R}^{-1}(\tau)(\tilde{y}(\tau)-\tilde{r}(\tau)),
    \end{aligned}
\end{equation}
where the terminal cost is defined by $S(\tau_f) = P_0^{-1}$ and $v(\tau_f) = P_0^{-1} m_0$. Consequently, we have
\begin{equation}\label{control_law_linearcase}
    u^{*}(\phi,\tau)=- \tilde{Q}(\tau) S(\tau) \phi + \tilde{Q}(\tau) v(\tau).
\end{equation}
The optimal states can be obtained by solving the following differential equation:
\begin{equation}\label{smoothingsol}
    \begin{aligned}
        \dot{\phi^{*}}(\tau)=\tilde{F}(\tau)\phi^{*}
        + \tilde{c}(t) + u^{*}(\tau),
    \end{aligned}
\end{equation}
where $u^{*}(\tau) = u^{*}(\phi^*(\tau),\tau)$ which can be rewritten as 
\begin{equation}
    \dot{\phi^{*}}(\tau)
    = \bar{F}(\tau)\phi^{*}(\tau)+\bar{c}(\tau),
\end{equation}
such that
\begin{equation}
\begin{aligned}
    \bar{F}(\tau)&= \tilde{F}(\tau)-\tilde{Q}(\tau)S(\tau),\\
    \bar{c}(\tau)&= \tilde{Q}(\tau)v(\tau) + \tilde{c}(\tau),
\end{aligned}
\end{equation}
with the initial condition $\phi^{*}(\tau_0) = S(\tau_0)^{-1} v(\tau_0)$.

The solution to the above differential equation can be written in terms of the transition function $\Phi(\tau,\tau_0)\in \mathbb{R}^{n_x\times n_x}$ and $\beta(\tau,\tau_0)\in \mathbb{R}^{n_x}$ such that \cite{sarkka2019applied} 
\begin{equation}
    \dot{\phi}(\tau)=\Phi(\tau,\tau_0)\phi(\tau_0)+\beta(\tau,\tau_0),
\end{equation}
where
\begin{equation}\label{odetransition}
    \begin{aligned}
    \frac{\partial \Phi(\tau,\tau_0)}{\partial \tau }&=\bar{F}(\tau)\Phi(\tau,\tau_0),\\
    \frac{\partial\beta(\tau,\tau_0)}{\partial \tau }&=\bar{F}(\tau)\beta(\tau,\tau_0)+\bar{c}(\tau),
    \end{aligned}
\end{equation}
with the initial conditions $\Phi(\tau,\tau_0)=I_{n_x}$ and $\beta(\tau,\tau_0)=0$. This reformulation makes the forward pass easier for parallel computing, which is discussed in Section \ref{otr_lqt}.

\subsection{Correspondence with Kalman-Bucy filtering and RTS smoothing}\label{sec:corr-kfs}

 The Riccati-type equations \eqref{2df} turn out to correspond to the information form of the Kalman--Bucy filter \cite{todorov2008general}, such that the information matrix (the inverse posterior covariance) is $S(\tau)$ and the information vector is $v(\tau)$. We make the change of variables $P(t) = S(t_f - t)^{-1}$ and $m(t) = S(t_f - t)^{-1} \, v(t_f - t)$, where $m(t)$ and $P(t)$ are mean and covariance of the filtering distribution, respectively. Then, putting back the original model matrices instead of the time-reversed ones then yields the Kalman--Bucy equations \cite{sarkka2019applied}
\begin{equation}
\begin{aligned}
\dot{m}(t) &= F(t)\,m(t) + c(t) + \\ & \qquad P(t)\,H(t)^{\top}R(t)^{-1}\big(y(t)-H(t)m(t)-r(t)\big),\\
\dot{P}(t) &= F(t)P(t) + P(t)F(t)^{\top} + Q(t)\\
  & \qquad - P(t)H(t)^{\top}R(t)^{-1}H(t)P(t),
\end{aligned}
\end{equation}
Furthermore, the initial conditions are  $P(t_0) = P_0$ and $m(t_0) = m_0$.

Reversing the time in \eqref{control_law_linearcase} and substituting $S(t_f - t) = P(t)^{-1}$ and $v(t_f - t) = P(t)^{-1} m(t)$ gives
\begin{equation}
    \bar{u}^{*}(x,t)= - Q(t) P(t)^{-1} x + Q(t) P(t)^{-1} m(t).
\end{equation}
Substituting to the time-reversed \eqref{smoothingsol} then gives
\begin{equation}
    \begin{aligned}
        \dot{x}^{*}(t) &= F(t)x^{*} (t) + c(t) - u^{*}(t) \\
        &= F(t)x^{*}(t) + c(t) + Q(t) P(t)^{-1} [x^*(t) - m(t)],
    \end{aligned}
\end{equation}
which can be recognized as the continuous-time RTS smoother mean equation \cite{sarkka2019applied}, where the terminal condition is $x^*(t_f) = m(t_f)$.

Overall, the backward equations in the continuous optimal-control formulation, up to time reversal and the information–covariance transformation, are equivalent to the Kalman–Bucy filter, while the forward equations correspond to the RTS smoother. This implies that parallelising the backward and forward passes of the continuous optimal control solver for the linear-affine problem will directly parallelise the Kalman–Bucy filter and RTS smoother in information form.

\section{Temporal parallelization of continuous-time MAP estimation}

After presenting the sequential continuous-time MAP estimation using the optimal-control approach, we now turn to parallel computation to reduce the computational cost to logarithmic in the number of steps. We start with a brief introduction to the parallel scan algorithm.

\subsection{Parallel computing with associative operators}
The parallel associative scans algorithms \cite{Blelloch:1989}, also known as the all-prefix-sum algorithms or scan algorithms, are used in this paper to parallelise the continuous-time optimal control solution for the MAP problem defined in Section~\ref{sec:control-map}.

A parallel associative scan is a parallel algorithm that computes all-prefix-sums for an associative operator $\otimes$. Suppose that we have a sequence of elements as $a_1,a_2,\ldots,a_T$ and we can define an associative operator $\otimes$ on these elements. Then, the all-prefix-sums operation computes the sequence $s_1,s_2, \ldots,s_T$ such that
\begin{equation}
    \begin{aligned}
        s_1 & = a_1,\\
        s_2 & = a_1 \otimes a_2,\\
        \vdots &\\
        s_T & = a_1 \otimes a_2 \otimes  \cdots \otimes a_T.
    \end{aligned}
\end{equation}
It should be noted that the associative property means that $(a\otimes b) \otimes c=(c \otimes a) \otimes b$ for any three elements $a,b,c$ and the defined operator $\otimes$. For instance, the sum, multiplication, and minimisation are associative operators.

The sequential computation of the all-prefix-sums operation has complexity $\mathcal{O}(T)$, even if we use GPUs. Therefore, a more efficient way is to use the parallel scan algorithm \cite{Blelloch:1989}, which applies parallelisation to compute the all-prefix-sums. This reduces the computational complexity, or more specifically, the span complexity, to $\mathcal{O}(\log T)$. This parallel method uses the up-sweep and down-sweep computations on a binary tree that splits the problem into independent sub-problems, which can be performed in parallel \cite{Blelloch:1989}.

Similarly, it is also possible to compute reversed all-prefix-sums, which are given by
\begin{equation}
    \begin{aligned}
        \bar{s}_1 & = a_1 \otimes a_2 \otimes  \cdots \otimes a_T,\\
        \bar{s}_2 & = a_2 \otimes  \cdots \otimes a_T,\\
        \vdots &\\
        \bar{s}_T & = a_T.
    \end{aligned}
\end{equation}
This can be done by reversing the sequence before and after applying the parallel associative scan operation. For a recent review and expirical comparison of different parallel associative scan algorithms in the discrete-time Kalman filtering and smoothing contexts, see \cite{sarkka2025perf}.

\subsection{Conditional value functions and combination rule}

To solve the MAP estimation problem in parallel, we adopt the parallel continuous optimal-control method presented in \cite{sarkka2024temporal}, which is based on parallel associative scans. The elements $a_1,a_2,\ldots,a_T$ in the parallel associative scan are now conditional value functions, which we define as follows.

\begin{defn} The conditional value function $V(\phi,s;z,\gamma)$ from $s\in[\tau_0,\tau_f]$ and $\gamma \in[\tau_0,\tau_f]$  is the cost of the optimal trajectory starting from $\phi(s)=\phi$ to $\phi(\gamma)=z$. Then the conditional value function when $z \in \mathcal{A} (\phi,s;\gamma)$ is as follows:
\begin{equation}\label{conditionalvf}
    V(\phi,s;z,\gamma)=\min_{u}\left\{\int_{s}^{\gamma}l(\phi(\tau),u(\tau))d\tau |\phi(\gamma)=z\right\},
\end{equation}
subject to the dynamic model
\begin{equation}
   \dot{\phi}(\tau)=\tilde{f}(\phi,\tau)+u(\tau),
\end{equation}
where $\phi(s)=\phi$ is given and $\mathcal{A} (\phi,s;\gamma)$ is a set, which includes reachable states at time $\gamma$ from the state $x(s)$. If the state $z \notin \mathcal{A} (\phi,s;\gamma)$, then the cost from the state $\phi$ and $z$ is infinity, $V(\phi,s;z,\gamma)=\infty$.
\end{defn}

We can combine two value functions, one defined on the interval $[s,\gamma]$ and the other on $[\gamma,\rho]$, to obtain the conditional value function from time $s$ to $\rho$ as follows. 

\begin{defn}. Suppose we have two functions $G_1, G_2: \mathbb{R}^{n_x} \times \mathbb{R}^{n_x} \rightarrow \mathbb{R}$. Define their combination as
\begin{equation}
    (G_1 \otimes G_2)(x,y) = \min_{z \in \mathbb{R}^{n_x}} \left( G_1(x,z) + G_2(z,y) \right).
\end{equation}
It is direct to check that this operation is associative. 
\end{defn}

Furthermore, the conditional value function from time $s$ to $\tau$, for any $s < \gamma < \tau$ and any $\phi, z \in \mathbb{R}^{n_x}$, satisfies \cite{sarkka2024temporal}
\begin{equation}\label{valuefuncion-assoc}
    V(\phi,s;z,\tau)=V(\phi,s;\cdot,\gamma) \otimes V(\cdot,\gamma;z,\tau).
\end{equation}

\subsection{Backward conditional HJB equation}

 The backward HJB equation for the conditional value function \cite{sarkka2024temporal} of the optimal control problem defined in Section~\ref{optimalcontrolproblem} is now given as
 \begin{equation}\label{backwardHJB}
     \begin{aligned}
         -&\partial_{s} V(\phi,s;z,\gamma)=\\&\min_u\left\{ l_s(\phi,u) + \partial_{\phi} V(\phi,s;z,\gamma)^{\top}( \tilde{f}(\phi,s)+u)\right\},
     \end{aligned}
 \end{equation}
which evolves the conditional value function from the final time $\gamma$ back to an earlier time $s$. The optimal control can then be solved as
\begin{equation}
     u^{*}(s;z,\gamma)=-\tilde{Q}(s)\partial_\phi V(\phi,s;z,\gamma).
\end{equation}
Substituting this into \eqref{backwardHJB} gives
 \begin{equation}\label{backwardconHJB}
     \begin{aligned}
    -\partial_s  &V(\phi,s;z,\gamma)
      =-\frac{1}{2}\partial_{\phi} V(\phi,s;z,\gamma)^{\top}\tilde{Q}(s)\partial_{\phi} V(\phi,s;z,\gamma)\\& 
      +\frac{1}{2}\left(\tilde{y}(s)-\tilde{h}(\phi,s)\right)^{\top}\tilde{R}(s)^{-1}\left(\tilde{y}(s)-\tilde{h}(\phi,s)\right)\\&
      +\partial
      _{\phi}J(\phi,s;z,l)^{\top}\tilde{f}(\phi,s)-\frac{1}{2}\nabla \cdot \tilde{f}(\phi,s).
     \end{aligned}
 \end{equation}
To complete the backward HJB formulation, we need a terminal boundary condition at the final time $\gamma$. Because the problem enforces a fixed terminal state $z$, the conditional value function must encode that only trajectories satisfying $\phi(\gamma) = z$ are feasible. Thus, the terminal condition is given as \cite{sarkka2024temporal}
\begin{equation} \label{eq:V_cond_boundary}
V(\phi,\gamma;z,\gamma) =
\begin{cases}
0, & \text{if } \phi = z, \\
\infty, & \text{if } \phi \neq z,
\end{cases}
\end{equation}
which assigns zero cost when the terminal constraint is satisfied and an infinite penalty otherwise.

\subsection{Parallel optimal control solution} \label{sec:paropt}

To construct a practically implementable algorithm, similarly to \cite{sarkka2024temporal},  the time interval $[\tau_0,\tau_f]$ must be divided into $T$ sub-intervals as $[\tau_0,\tau_1],\ldots,[\tau_{T-1},\tau_T]$, where $\tau_T=\tau_f$. Then the value functions can be recovered by defining associative elements $a_i$ for $i=0,\ldots,T$, where they are as follows:
 \begin{equation}
     \begin{aligned}
         a_i&=V(\cdot,\tau_i;\cdot,\tau_{i+1}), \quad \textit{for} \quad  i=1,2,\cdots,T-1,\\
         a_T&=V(\cdot,\tau_T),
     \end{aligned}
 \end{equation}
 which implies that
 \begin{equation}\label{valueparallel}
     \begin{aligned}
         a_0 \otimes a_1 \otimes \cdots \otimes a_k & = V(\cdot,\tau_0;\cdot,\tau_{k+1}), \\
         a_k \otimes a_{k+1} \otimes \cdots \otimes a_T &=V(\cdot,\tau_k),
     \end{aligned}
 \end{equation}
 where the associative operator $\otimes$ is defined in \eqref{valuefuncion-assoc}. For each sub-interval, the element is initiated in parallel for each $i$. Specifically, the element $a_T$ is $V(\cdot,\tau_T)=-\log(p(\phi_{t_{f}}))$, and other elements $a_i$ for $i=0,1,2,\ldots,T-1$ can be obtained by solving the backward HJB equation \eqref{backwardconHJB} in parallel for each interval $\tau \in [\tau_i, \tau_{i+1}]$. 
 
The value functions at the times $\tau_i$ for $i=0,1,2,\ldots,T-1$ can then be obtained by applying the parallel associative scans to the elements $a_i$ with the operator $\otimes$. Finally, the backward HJB equation \eqref{backwardconHJB} can be used to obtain the value function at any desirable time. The resulting algorithm is parallel algorithm which corresponds to the forward pass of the Viterbi algorithm in discrete-time case.

 \subsection{Optimal trajectory and MAP estimate recovery}\label{otr}
 
In the previous section, we derived a parallel computational method for the value functions using associative scan operations. We now turn to computing the backward optimal trajectory $\phi^{*}(\tau)$, which is the solution to \eqref{controldynamic}. 

First note that given the value functions, we can compute the optimal controls via \eqref{optimalcontrollaw} fully in parallel. We can now leverage the first method described in \cite{sarkka2024temporal}, to find the solution $\phi^*(\tau)$ at an arbitrary time $\tau$. Let $\alpha_{i}(\phi)$ denote the function which gives the solution to \eqref{controllaw} at time $\tau_i$ given initial condition $\phi(\tau_{i-1}) = \phi$ at time $\tau_{i-1}$. We can then form the full solution at time step $\tau_i$ as
\begin{equation}\label{associativerecovery}
    \begin{aligned}
        \phi^{*}(\tau_i)=(\alpha_i \circ \alpha_{i-1} \circ \cdots \circ \alpha_{1})(\phi^*(\tau_0)),
    \end{aligned}
\end{equation}
where $\circ$ denotes the function composition and $\phi^{*}(\tau_0) = \arg\min_{\phi}V(\phi,\tau_0)$. The composition is an associative operator and therefore \eqref{associativerecovery} can be implemented in parallel using a parallel associative scan. Given the values at the time steps $\tau_i$, we can sequentially solve \eqref{controllaw}  at each time interval to give the solution $\phi^*(\tau)$ at an arbitrary time. The final MAP trajectory can then be recovered as $x^*(t) = \phi^{*}(t_f - t)$.

Instead of propagating the state forward through the system dynamics, there is also a second method to recover the trajectory from the optimal value function given in \cite{sarkka2024temporal}. However, as in our case we do not have an initial condition, the procedure needs to be change a bit. The direct application of the second method in \cite{sarkka2024temporal} gives
\begin{equation}
    \bar{\phi}^{*}(\tau) = \arg\min_{\phi}\left\{V(\phi(\tau_0),\tau_0,\phi,\tau)+V(\phi,\tau)\right\},
\end{equation} 
which is conditioned on $\phi(\tau_0)$ however, we can form the corresponding solution for the free initial condition case simply by minimizing over the initial condition as well, which gives
\begin{equation}
    \phi^{*}(\tau) = \arg\min_{\phi}\left\{ \min_{\phi(\tau_0)} V(\phi(\tau_0),\tau_0,\phi,\tau)+V(\phi,\tau)\right\}.
\label{eq:v-comb}
\end{equation} 

In practice, the forward value functions $V(\phi(\tau_0),\tau_0,\phi,\tau)$ can be computed for times $\tau = \tau_i$ by performing a forward associative scan for elements $a_i$ that we defined in Section~\ref{sec:paropt}. However, to compute them from arbitrary times, instead of using the backward HJB in \eqref{backwardconHJB}, it is computationally more efficient to use the forward HJB equation which is given as \cite{sarkka2024temporal}
\begin{equation}\label{eq:forward_conditional_HJB}
\begin{split}
&\frac{\partial V(x,s;z,\tau)}{\partial \tau}\\
&= \min_{u} \left\{
\ell(z,u,\tau)
- \left( \frac{\partial V(x,s;z,\tau)}{\partial z} \right)^\top f(z,u,\tau)
\right\}.
\end{split}
\end{equation}

Above \eqref{associativerecovery} now corresponds to a parallel version of the classical forward-backward Viterbi algorithm in discrete time, and \eqref{eq:v-comb} to the max-product formulation. In the linear-affine case, these correspond to the RTS smoother and the two-filter smoother, respectively.

\section{Temporal parallelization of continuous-time Kalman-Bucy filtering and smoothing}
In this section, we focus on the linear–affine case of the MAP estimation problem. As discussed in Sections~\ref{LQT} and \ref{sec:corr-kfs}, the continuous-time LQT problem corresponding to minimisation of the Onsager--Machlup equation in the linear-affine case is equivalent to the Kalman–Bucy filter and the continuous-time RTS smoother. Therefore, parallelizing the LQT is equivalent to parallelizing the Kalman–Bucy filter and the continuous-time RTS smoother. 

\subsection{Combination rule for the linear affine case}
As shown in \cite{sarkka2024temporal}, the conditional value function for the affine-linear model in \eqref{affine} is a quadratic form. However, on a time interval of vanishing length, it becomes singular, and therefore, in \cite{sarkka2024temporal}, it was formulated in a dual form. However, the parameterisation of the conditional value function essentially has the form
\begin{equation}\label{conditionalvalue_linear}
 \begin{aligned}
    V(\phi,s;z,\gamma) &= \text{constant} + \frac{1}{2}\phi^{\top} J(s,\gamma)\phi-\phi^{\top}\eta(s,\gamma)\\
         &+\frac{1}{2}\left(z-A(s,\gamma)\phi-b(s,\gamma)\right)^{\top}C(s,\gamma)^{-1}\\
        & \quad \times  \left(z-A(s,\gamma)\phi-b(s,\gamma)\right),
 \end{aligned}
\end{equation}
where the defining parameters are $A$, $b$, $C$, $J$, and $\eta$. This explicit form is strictly valid only when $C$ is invertible, hence the dual formulation in \cite{sarkka2024temporal}. 
 
Now, suppose that the conditional value functions $V(\phi,s;\cdot,\gamma)$ and
$V(\cdot,\gamma;z,t)$ are expressed in the form \eqref{conditionalvalue_linear}, and that the intermediate time satisfies $s < \gamma < t$. Then the parameters $A$, $b$, $C$, $J$, and $\eta$ defining the quadratic value function  $V(\phi,s;z,t) = V(\phi,s;\cdot,\gamma) \otimes V(\cdot,\gamma;z,t)$ can be obtained using \eqref{valuefuncion-assoc} which reduces to
\begin{equation}
    \begin{aligned}
A(s,t) &= A(\gamma,t)\big(I_{n_x} + C(s,\gamma) J(\gamma,t)\big)^{-1} A(s,\gamma), \\
b(s,t) &= A(\gamma,t)\big(I_{n_x} + C(s,\gamma) J(\gamma,t)\big)^{-1}  \\
&\quad \times \big(b(s,\gamma) + C(s,\gamma)\eta(\gamma,t)\big) + b(\gamma,t), \\
C(s,t) &= A(\gamma,t)\big(I_{n_x} + C(s,\gamma) J(\gamma,t)\big)^{-1} C(s,\gamma) A(\gamma,t)^\top \\
&\quad + C(\gamma,t), \\
\eta(s,t) &= A(s,\gamma)^\top \big(I_{n_x} + J(\gamma,t) C(s,\gamma)\big)^{-1}  \\
&\quad \times \big(\eta(\gamma,t) - J(\gamma,t)b(s,\gamma)\big) + \eta(s,\gamma), \\
J(s,t) &= A(s,\gamma)^\top \big(I_{n_x} + J(\gamma,t) C(s,\gamma)\big)^{-1} J(\gamma,t) A(s,\gamma)  \\
&\quad + J(s,\gamma),
\end{aligned}
\end{equation}
where $I_{n_x}$ is an identity matrix. 
Since $C(s,\gamma)$ and $J(\gamma,t)$ are symmetric positive semi-definite matrices, the matrices $I_{n_x}+C(s,\gamma)J(\gamma,t)$ and $I_{n_x}+J(s,\gamma)C(\gamma,t)$ are invertible \cite{sarkka2024temporal}. This operation is associative.

However, we need to determine these parameters for non-zero intervals. For that purpose, we need to solve the backward conditional HJB equation for the special LQT case, which is discussed in the next section.

\subsection{Backward conditional HJB equation for the linear-affine case}

In this section, we discuss the backward HJB equation for the conditional value functions of linear-affine models. In this case, the conditional HJB reduces to backward differential equations for the parameters  $A,b,C,\eta,J$ which are given as follows \cite{sarkka2024temporal}:
\begin{equation}\label{backwardLQT}
    \begin{aligned}
        \partial_s A(s,\gamma)&=-A(s,\gamma)\tilde{Q}(s)J^{\top}(s,\gamma)-A(s,\gamma)\tilde{F}(s)\\
        \partial_s b(s,\gamma)&=-A(s,\gamma)\tilde{Q}(s)\eta(s)-A(s,\gamma)\tilde{c}(s)\\
         \partial_s C(s,\gamma)&=-A(s,\gamma)\tilde{Q}(s)A^{\top}(s,\gamma)\\
          \partial_s \eta(s,\gamma)&=J(s,\gamma)\tilde{Q}(s)\eta(s,\gamma)-\tilde{F}^{\top}(s)\eta(s,\gamma)\\
          &-\tilde{H}(s)^{\top}\tilde{R}^{-1}(s)(\tilde{y}(s)-\tilde{r}(s)) + J(s,\gamma)\tilde{c}(s)\\
           \partial_s J(s,\gamma)&=J(s,\gamma)\tilde{Q}(s)J^{\top}(s,\gamma)-J(s,\gamma)\tilde{F}(s) \\
           & \quad -\tilde{F}(s)^{\top}J(s,\gamma)-\tilde{H}(s)^{\top}\tilde{R}^{-1}(s)\tilde{H}(s). 
    \end{aligned}
\end{equation}
The boundary conditions corresponding to \eqref{eq:V_cond_boundary} are $A(\tau,\tau)=I_{n_x}$, $b(\tau,\tau)=0$, $C(\tau,\tau)=0$, $\eta(\tau,\tau)=0$, and $J(\tau,\tau)=0$ for all $\tau\in[\tau_0,\tau_f]$. 

A useful thing to notice is that $J(s,\gamma)$ and $\eta(s,\gamma)$ take the role of Riccati ODEs parameters $S(\cdot)$ and $v(\cdot)$ in \eqref{2df}. If we set $J(s,\gamma)=P(s,\gamma)^{-1}$ and $\eta(s,\gamma)=P(s,\gamma)^{-1}m(s,\gamma)$, then we have 
\begin{equation}
\begin{aligned}
\partial_s &m(s,\gamma) = \tilde{F}(s) m(s,\gamma) + \tilde{c}(s)
\\& - P(s,\gamma) \tilde{H}(s)^{\top} \tilde{R}^{-1}(s) \left( \tilde{y}(s)-\tilde{H}(s) m(s,\gamma)- \tilde{r}(s) \right), \\
\partial_s &P(s,\gamma) = P(s,\gamma) \tilde{F}(s) + \tilde{F}^\top(s) P(s,\gamma) \\ &
+ P(s,\gamma) \tilde{H}^\top(s) \tilde{R}^{-1}(s) \tilde{H}(s) P(s,\gamma) - \tilde{Q}(s),
\end{aligned}
\end{equation}
which are essentially the Kalman-Bucy filter equations in backward time. Henceforth, in principle, it would be possible to replace $\eta$ and $J$ with their mean and covariance counterparts. However, by doing so, we run into a problem with representing the boundary conditions, which would require an infinite covariance.

\subsection{Optimal trajectory and MAP estimate recovery}\label{otr_lqt}
In Section \ref{otr}, we discussed how the optimal trajectory can be obtained using two different methods. Here, we show how to specialise them to the linear–affine state-space model. The first method corresponds to computing $\Phi(\tau,\tau_{i-1})$ and $\beta(\tau,\tau_{i-1})$ 
by solving ODEs in \eqref{odetransition} in parallel for each sub-interval. Then assume that over $[s,\tau]$, we already know 
$\Phi(\tau,s)$ and $\beta(\tau,s)$,  
and over $[\tau,\gamma]$ we have 
$\Phi(t,\tau)$ and $\beta(t,\tau)$.  
Then the quantities for the entire interval $[s,\gamma]$ follow from the combination rule:
\begin{align}
    \Phi(\gamma,s) &= \Phi(\gamma,\tau)\,\Phi(\tau,s), \label{combination_forward_phi}\\
    \beta(\gamma,s) &= \Phi(\gamma,\tau)\,\beta(\tau,s) + \beta(\gamma,\tau). \label{combination_forward_beta}
\end{align}
which defines an associative operator. As a result, we can compute $\Phi(\tau_i,\tau_{i-1})$ and $\beta(\tau_i,\tau_{i-1})$ for all $i$ and use a parallel scan to combine them using \eqref{combination_forward_phi} and \eqref{combination_forward_beta}, yielding  
$\Phi(\tau_i,\tau_0)$ and $\beta(\tau_i,\tau_0)$ for every $i$. 

The associative scan together with solving the ODEs at sub-interval then yields the quantities $\Phi(\tau,\tau_0)$ and $\beta(\tau,\tau_0)$,
which can be used to solve the optimal state as
\begin{equation}
\phi^*(\tau) = \Phi(\tau,\tau_0)\phi^*(\tau_0) + \beta(\tau,\tau_0).
\label{eq:42}
\end{equation}

The second method, as mentioned in Section \ref{otr}, is equivalent to the two-filter smoother. In this case, the optimal state trajectory can be obtained from the value function and the conditional value function for the LQT directly as
\begin{equation}\label{twofilter}
    \begin{aligned}
        \phi^{*}(\tau)&=(I_{n_x} + \bar{C}(\tau_0,\tau) S(\tau))^{-1}\\& \times (\bar{b}(\tau_0,\tau) + \bar{C}(\tau_0,\tau)v(\tau)),
    \end{aligned}
\end{equation}
where $S$ and $v$ are computed by the backward pass, and parameters $\bar{C}$, and $\bar{b}$ are the parameters of $\min_{\phi(\tau_0)} V(\phi(\tau_0),\tau_0,\phi,\tau)$, which can be obtained as parameters of $e \otimes V(\phi(\tau_0),\tau_0,\phi,\tau)$, where $e$ is the constant conditional value function that can be parametrised as
\begin{equation}
\begin{aligned}
    A_e&= 0_{n_x\times n_x}, \quad b_e= 0_{n_x \times 1}, \quad C_e=\kappa I_{n_x}, \\
    \eta_e&= 0_{n_x\times1}, \quad J_e = 0_{n_x\times n_x},
\end{aligned}
\end{equation}
where $\kappa \rightarrow \infty$. In practice, we can implement this computation by computing the prefix sums for a sequence which is obtained by replacing the first element $a_0$ in the sequence $a_0 = V(\phi,\tau_{0};z,\tau_{1})$, $a_1 = V(\phi,\tau_{1};z,\tau_{2})$, $a_2 = V(\phi,\tau_{3};z,\tau_{3})$, \ldots with $\bar{a}_0 = e \otimes a_0$, and finally inserting $e$ in front of the prefix sequence. If $a_0$ has the parameters $(A_0,b_0,C_0,\eta_0,J_0)$, then $\bar{a}_0$ has the parameters
\begin{equation}
\begin{aligned}
  \bar{A}_0 &= 0, \\
  \bar{b}_0 &= A_0 \, J_0^{-1} \, \eta_0, \\
  \bar{C}_0 &= A_0 \, J_0^{-1} \, A_0^\top + C_0, \\
  \bar{\eta}_0 &= \eta_0, \\
  \bar{J}_0 &= J_0.
\end{aligned}
\end{equation}

To recover $\bar{b}(\tau_0,\tau)$ and $\bar{C}(\tau_0,\tau)$ for arbitrary times $\tau$ (not just $\tau_i$), it is then convenient to use the forward HJB equation \eqref{eq:forward_conditional_HJB} which in the linear-affine case reduces to
\begin{equation}
\begin{split}
\partial\tau A\left(s,\tau\right)
&=-C\left(s,\tau\right)\tilde{H}\left(\tau\right)^{\top}\tilde{R}^{-1}\left(\tau\right)\tilde{H}\left(\tau\right)A\left(s,\tau\right) \\
&\quad+\tilde{F}\left(\tau\right)A\left(s,\tau\right), \\
\partial_\tau b\left(s,\tau\right)
&=C\left(s,\tau\right)\tilde{H}\left(\tau\right)^{\top}\tilde{R}^{-1}\left(\tau\right)[\tilde{y}\left(\tau\right) - \tilde{r}\left(\tau\right)] \\
&\quad + \tilde{F}\left(\tau\right) b\left(s,\tau\right) +\tilde{c}\left(\tau\right) \\
&\quad-C\left(s,\tau\right)\tilde{H}\left(\tau\right)^{\top}\tilde{R}^{-1}\left(\tau\right)\tilde{H}\left(\tau\right)b\left(s,\tau\right), \\
\partial_\tau C\left(s,\tau\right)
&=-C\left(s,\tau\right)\tilde{H}\left(\tau\right)^{\top}\tilde{R}^{-1}\left(\tau\right)\tilde{H}\left(\tau\right)C\left(s,\tau\right) \\
&\quad+\tilde{Q}\left(\tau\right)+\tilde{F}\left(\tau\right)C\left(s,\tau\right) \\
&\quad+C\left(s,\tau\right)\tilde{F}\left(\tau\right)^{\top}, \\
\partial_\tau \eta\left(s,\tau\right)
&=A\left(s,\tau\right)^{\top}\tilde{H}\left(\tau\right)^{\top}\tilde{R}^{-1}\left(\tau\right) [\tilde{y}\left(\tau\right) - \tilde{r}\left(\tau\right)] \\
&\quad-A\left(s,\tau\right)^{\top}\tilde{H}\left(\tau\right)^{\top}\tilde{R}^{-1}\left(\tau\right)\tilde{H}\left(\tau\right)b\left(s,\tau\right), \\
\partial_\tau J\left(s,\tau\right)
&=A\left(s,\tau\right)^{\top}\tilde{H}\left(\tau\right)^{\top}\tilde{R}^{-1}\left(\tau\right)\tilde{H}\left(\tau\right)A\left(s,\tau\right),
\end{split}
\label{eq:Forward_HJB_LQT}
\end{equation}
from which we, in practice, only need to first three equations.

By reversing the time, we can obtain the MAP state trajectory $x^*(t) = \phi^*(t_f - t)$. The procedure based on \eqref{eq:42} now is the parallel continuous-time RTS smoother, and the latter procedure \eqref{twofilter} is the parallel two-filter smoother. In both of the methods, $P = S^{-1}$ and $m = S^{-1} \, v$ correspond to Kalman--Bucy filter's mean and covariance. In the latter method, $\bar{C}$ and $\bar{b}$ correspond to the covariance and mean of the backward filter, respectively.

\subsection{Non-linear state-space models}\label{non-linear-section}
We can handle nonlinear models using iterative linearisation: we first linearise the estimation problem \eqref{statespace} about a nominal trajectory, then solve the resulting linear-affine problem using the model from the previous sections. We then linearise again using the solution to the linear-affine problem and so on. This method is equivalent to a continuous-time iterated extended Kalman smoother\cite{Sarkka+Svensson:2023}, and we can implement it in parallel, either in the RTS smoother form or in the TF smoother form.

\section{Experimental results}

In this section, we present experimental results on the runtime of the proposed methods on a graphics processing unit (GPU). We compare the parallel methods with their sequential counterparts. All the experiments were run on an NVIDIA Tesla A100, 80-gigabyte GPU. The experimental results indicate that the method gives us a computational advantage for computing the MAP solution. In the experiments, we consider both linear and non-linear models.

\subsection{Linear state-space model}
In this section, we consider the partially  observed Wiener velocity model \cite{sarkka2019applied} which has the form:
\begin{equation}
\begin{aligned}
    \dot{x}(t)&=
    Fx(t)+Lw(t),\\
    y(t)&=Hx(t)+\nu(t),
\end{aligned}
\label{wiener-vel}
\end{equation}
where the state $x(t)$ includes the 2D positions and velocities $\begin{pmatrix}p^{x}_t & p^y_t & v^x_t & v^y_t\end{pmatrix}^{\top}$, and we assume that the observations consist of the positions corrupted by measurement noise. Here, $w(t)$ and $\nu(t)$ are white Gaussian noise. The matrices in the model are:
\begin{equation}
    F=\begin{pmatrix}
        0_{2\times 2} & I_{2\times 2}\\
        0_{2\times 2}  & 0_{2 \times 2}
    \end{pmatrix}, 
    H=\begin{pmatrix}
        I_{2 \times 2} & 0_{2 \times 2}
    \end{pmatrix}, 
    L=\begin{pmatrix}
        0_{2 \times 2} \\ I_{2 \times 2}\\
    \end{pmatrix}.
\end{equation}
The time span $t \in [0,5]$ and the spectral densities and initial parameters are as follows:
\begin{equation}
    \begin{aligned}
    W&=4\, I_{2\times 2}, \quad R=10^{-2}\,I_{2\times2}\\
    m_0&=\begin{pmatrix}5 & 5 & 0 & 0\end{pmatrix}^{\top}, \quad P_0=10^{-2} \, I_{2 \times 2}.
    \end{aligned}
\end{equation}
 Sequential and parallel versions of the RTS and TF smoothers were implemented to compute $S(\tau)$ and $v(\tau)$ in the backward pass and $\phi(\tau)$ in the forward pass, with the hidden state $x(t)$ obtained by reversing time. In the parallel methods, the time interval is divided into $T$ blocks, each with $n=10$ substeps. The backward and forward passes within each block are solved using Euler’s method, and the results are combined through associative scans, yielding a computational complexity of $\mathcal{O}(\log T)$. In contrast, the sequential version performs Euler integration for all $nT$ steps. 
 
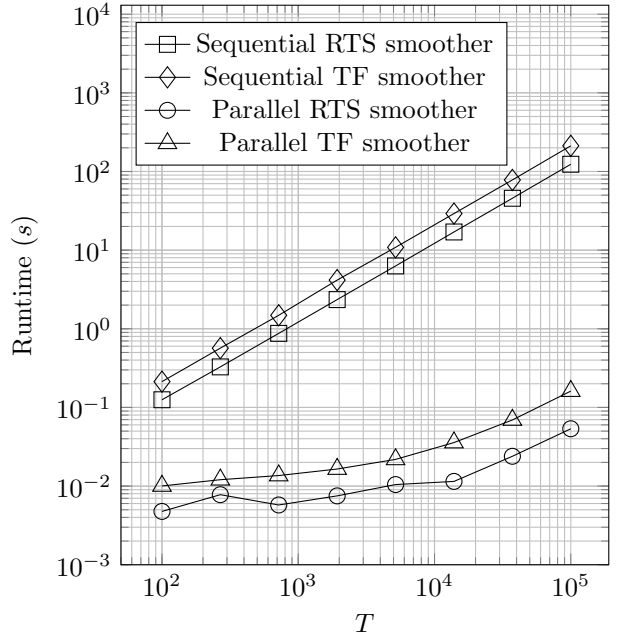
\begin{figure}     
     \centering
     \input{figures/linear_runtime}
     \caption{The runtime results for the sequential and parallel continuous-time MAP algorithms run on GPU for the Wiener velocity model \eqref{wiener-vel}.}
     \label{fig:seqvspar}
 \end{figure}

 Fig.~\ref{fig:seqvspar} presents the mean runtime over five iterations for both the sequential and parallel algorithms executed on the GPU. The results show that the parallel method for computing the MAP estimate is faster. As shown in the figure, the two-filter smoother takes longer to run than the RTS smoother, but still outperforms the sequential smoothers. 
 \subsection{Non-linear state-space model}
 In this section, we consider the continuous-time coordinated turn state-space model \cite{Sarkka+Svensson:2023}, which is a nonlinear model of a target taking a turn with a randomly varying turn rate. In Cartesian coordinates, the model can be written as follows:
 \begin{equation}
     \begin{aligned}
         \dot{x}(t)&=f(x(t))+Lw(t),\\
              y(t)&=h(x(t))+\nu(t),
     \end{aligned}
 \end{equation}
 where the state $x(t)=\begin{pmatrix}\xi_t &\zeta_t&\dot{\xi}_t&\dot{\zeta}_t&\omega_t\end{pmatrix}^{\top}$, and $w(t)$ and $\nu(t)$ are white Gaussian noise processes. Furthermore, the functions $f$ and $h$ are as follows:
 \begin{equation}
     \begin{aligned}
f(x(t))&=\begin{pmatrix}\dot{\xi}_t&\dot{\zeta}_t&-\omega_t \dot{\zeta}_t&\omega_t \dot{\xi}_t& 0 \end{pmatrix}^{\top},\\
         h(x(t))&=\begin{pmatrix}\sqrt{\xi_t^2+\zeta_t^2} & \arctan(\frac{\zeta_t}{\xi_t})\end{pmatrix}^{\top},
     \end{aligned}
     \label{ct-model}
 \end{equation}
 and the diffusion matrix is
\begin{equation}
     L=\begin{pmatrix}
       0 & 0 & 0 \\
       0 & 0 & 0 \\
       \sigma_{v_{\xi}} & 0 & 0\\
       0 & \sigma_{v_{\zeta}} & 0\\
       0 & 0 & \sigma_w
       \end{pmatrix},
 \end{equation}
 where $\sigma_{v_{\xi}}=\sigma_{v_{\zeta}}=5 \times 10^{-4}$ and $\sigma_{\omega}=0.02$. The spectral density and initial parameters for the problem are
 \begin{equation}
     \begin{aligned}
       W&=I_{3\times 3}, \\
       R&=\mathrm{diag}(5 \times 10^{-3},10^{-3}),\\ 
       m_0&=\begin{pmatrix}
           5 & 5 & 0 & 0.3 & 0
       \end{pmatrix}^{\top}, \\ P_0&=\mathrm{diag}(0.01,0.01,0.01,0.01,0.04).
     \end{aligned}
 \end{equation}
 
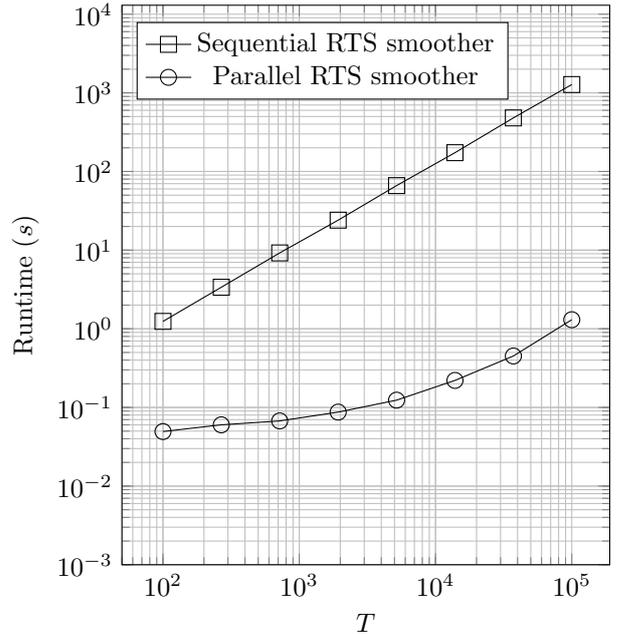
\begin{figure}     
     \centering
     \input{figures/nonlinear_runtime}
     \caption{The GPU runtime results for coordinated turn model \eqref{ct-model} using sequential and parallel iterated MAP algorithms.}
     \label{fig:seqvspar_nonlinear}
 \end{figure}

We implemented an iterated linearisation-based MAP estimation method for this problem. The linear-affine problem at each iteration was solved using the continuous-time sequential and parallel RTS smoothers. The number of iterations in the method was $5$. As shown in Fig.~\ref{fig:seqvspar_nonlinear}, the sequential method is significantly slower than the parallel method. In the figure, we also see that although the parallel method initially scales logarithmically with the number of blocks, it falls back to linear scaling when the number of cores runs out. A similar behavior can be observed in the linear state-space model in the previous section. Furthermore, since the TF smoother is computationally more expensive than the RTS smoother, as demonstrated in the linear case, we excluded it from the nonlinear state-space model experiments.

\section{Conclusions}
In this paper, we derived a parallel computational method for continuous-time MAP estimation in nonlinear stochastic state-space models. We reformulate the problem as a continuous-time optimal control problem based on the Onsager–Machlup cost functional, which allows us to leverage existing parallel optimal control techniques for MAP estimation. For the linear–Gaussian case, the formulation reduces to a parallel Kalman--Bucy filter, the continuous-time parallel RTS smoother, and the continuous-time parallel two-filter smoother. We then adapt the parallel approach to nonlinear systems using iterative Taylor linearisation. Finally, our GPU-based experiments show that the proposed parallel algorithms achieve a significant speed-up compared to their sequential counterparts. 

It is worth noting that it would be possible to derive the continuous-time Kalman filters and smoothers given in this paper also by taking the continuous-time limits of the corresponding discrete-time filters and smoothers \cite{Sarkka:2021, sarkka2025perf}. However, this approach would be less principled than the MAP estimation approach that we use in this article. An alternative approach would be to directly derive parallel versions of the continuous-time Kushner equations \cite{Kushner:1964, Bucy:1965} as well as the nonlinear smoothing equations \cite{Leondes+Peller+Stear:1970, Anderson:1972}, but this path of research is left as future work.

\begin{ack}                               
Supported by the Finnish Doctoral Program Network in Artificial Intelligence (AI-DOC). The authors thank Casian Iacob for his help.
\end{ack}

\bibliographystyle{plain}        
\bibliography{continuousviterbi}           
\appendix

\end{document}

%% file: figures/linear_runtime.tex
\begin{center}
 \begin{tikzpicture}
\begin{axis}[
width=8cm,    
height=9cm, 
legend pos=north west, 
    xmin=50, xmax=190000,
    ymin=0.001, ymax=13000,
    xmode=log, 
    ymode=log, 
    xlabel={$T$},
    ylabel={Runtime ($s$)},
    grid=both
]
\addplot[color=black, mark=square,mark size=3pt ] table [x index=0, y index=2, col sep=comma, header=false] {files/data.csv};
\addlegendentry{Sequential RTS smoother}
\addplot[color=black, mark=diamond,mark size=4pt] table [x index=0, y index=4, col sep=comma, header=false] {files/data.csv};
\addlegendentry{Sequential TF smoother}
\addplot[color=black, mark=o,mark size=3pt] table [x index=0, y index=1, col sep=comma, header=false] {files/data.csv};
\addlegendentry{Parallel RTS smoother}
\addplot[color=black, mark=triangle,mark size=4pt] table [x index=0, y index=3, col sep=comma, header=false] {files/data.csv};
\addlegendentry{Parallel TF smoother}

\end{axis}
\end{tikzpicture}
\end{center}

%% file: figures/nonlinear_runtime.tex
\begin{center}
 \begin{tikzpicture}
\begin{axis}[
width=8cm,    
height=9cm, 
legend pos=north west, 
    xmin=50, xmax=190000,
    ymin=0.001, ymax=13000,
    xmode=log, 
    ymode=log, 
    xlabel={$T$},
    ylabel={Runtime ($s$)},
    grid=both
]
\addplot[color=black, mark=square,mark size=3pt ] table [x index=0, y index=1, col sep=comma, header=false] {files/data2.csv};
\addlegendentry{Sequential RTS smoother}
\addplot[color=black, mark=o,mark size=3pt] table [x index=0, y index=2, col sep=comma, header=false] {files/data2.csv};
\addlegendentry{Parallel RTS smoother}

\end{axis}
\end{tikzpicture}
\end{center}

%% file: continuousviterbi.bib
@article{aihara2002mortensen,
  title={On the {M}ortensen equation for maximum likelihood state estimation},
  author={Aihara, Shin ichi and Bagchi, Arunabha},
  journal={IEEE Transactions on Automatic Control},
  volume={44},
  number={10},
  pages={1955--1961},
  year={1999},
  publisher={IEEE}
}

@article{Anderson:1972,
       title = {Fixed interval smoothing for nonlinear continuous time systems},
      author = {Anderson, Brian D. O.},
     journal = {Information and Control},
      volume = {20},
      number = {3},
       pages = {294--300},
        year = {1972},
   publisher = {Academic Press}
}

@article{bell1994iterated,
  title={The iterated {K}alman smoother as a Gauss--Newton method},
  author={Bell, Bradley M},
  journal={SIAM Journal on Optimization},
  volume={4},
  number={3},
  pages={626--636},
  year={1994},
  publisher={SIAM}
}

@book{Bellman:1957,
  title = {Dynamic Programming},
  author = {Bellman, Richard },
  publisher = {Princeton University Press},
  year = {1957},
}

@article{Blelloch:1989,
  author={Blelloch, Guy E.},
  journal={IEEE Transactions on Computers},
  title={Scans as primitive parallel operations},
  year={1989},
  volume={38},
  number={11},
  pages={1526--1538},
}

@article{Bucy:1965,
      author = {Bucy, R. S.},
       title = {Nonlinear filtering theory},
     journal = {IEEE Transactions on Automatic Control},
      volume = {10},
      number = {2},
        year = {1965},
       pages = {198--198}
}

@inproceedings{bryson1963,
  title={Smoothing for Linear and Nonlinear Dynamic Systems},
  author={Bryson, Arthur E. and Malcolm Frazier},
  booktitle={Proc. Optimum System Synthesis Conference},
  pages={353--364},
  year={1963}
}

@article{dutra2017joint,
  title={Joint maximum a posteriori state path and parameter estimation in stochastic differential equations},
  author={Dutra, Dimas Abreu Archanjo and Teixeira, Bruno Ot{\'a}vio Soares and Aguirre, Luis Antonio},
  journal={Automatica},
  volume={81},
  pages={403--408},
  year={2017},
  publisher={Elsevier}
}

@article{Dutra:2014,
       title = {Maximum a posteriori state path estimation: {D}iscretization limits and their interpretation},
      author = {Dutra, Dimas Abreu and Teixeira, Bruno Ot{\'a}vio Soares and Aguirre, Luis Antonio},
     journal = {Automatica},
      volume = {50},
       pages = {1360--1368},
        year = {2014},
   publisher = {Elsevier}
}

@article{hassan2021temporal,
  title={Temporal parallelization of inference in hidden Markov models},
  author={Hassan, Syeda Sakira and S{\"a}rkk{\"a}, Simo and Garc{\'\i}a-Fern{\'a}ndez, {\'A}ngel F.},
  journal={IEEE Transactions on Signal Processing},
  volume={69},
  pages={4875--4887},
  year={2021},
  publisher={IEEE}
}

@book{Jazwinski:1970,
      author = {Jazwinski, Andrew H.},
       title = {Stochastic Processes and Filtering Theory},
   publisher = {Academic Press},
     address = {New York},
        year = {1970}
}

@book{kirk2004optimal,
  title={Optimal control theory: an introduction},
  author={Kirk, Donald E},
  year={2004},
  publisher={Courier Corporation}
}

@article{Kushner:1964,
      author = {Kushner, Harold J.},
       title = {On the differential equations satisfied by conditional probability densities of {M}arkov processes, with applications},
     journal = {Journal of the Society for Industrial and Applied Mathematics, Series A: Control},
      volume = {2},
      number = {1},
       pages = {106--119},
        year = {1964}
}

@article{Larson+Peschon:1966,
  title={A dynamic programming approach to trajectory estimation},
  author={Larson, R. E. and Peschon, J.},
  journal={IEEE Transactions on Automatic Control},
  volume={11},
  number={3},
  pages={537--540},
  year={1966},
}

@article{Leondes+Peller+Stear:1970,
      author = {Leondes, Cornelius T. and  Peller, John B. and Stear, Edwin B. },
       title = {Nonlinear smoothing theory},
     journal = {IEEE Transactions on Systems Science and Cybernetics},
      volume = {6},
      number = {1},
       pages = {63--71},
       month = {January},
        year = {1970}
}

@article{liu2024onsager,
  title={The {O}nsager--{M}achlup action functional for degenerate stochastic differential equations in a class of norms},
  author={Liu, Shanqi and Gao, Hongjun},
  journal={Statistics \& Probability Letters},
  volume={206},
  year={2023},
  publisher={Elsevier}
}

@article{Omura:1969,
  title={On the {V}iterbi decoding algorithm},
  author={Omura, Jim},
  journal={IEEE Transactions on Information Theory},
  volume={15},
  number={1},
  pages={177--179},
  year={1969},
}

@book{sarkka2019applied,
  title={Applied Stochastic Differential Equations},
  author={S{\"a}rkk{\"a}, Simo and Solin, Arno},
  year={2019},
  publisher={Cambridge University Press}
}

@article{Sarkka:2021,
  title={Temporal Parallelization of {B}ayesian Smoothers},
  author={S{\"a}rkk{\"a}, Simo and Garc{\'\i}a-Fern{\'a}ndez, {\'A}ngel F.},
  journal={IEEE Transactions on Automatic Control},
  year={2021},
  volume={66},
  issue={1},
  pages={299--306},
}

@Article{Sarkka:2023,
  author={S{\"a}rkk{\"a}, Simo and Garc{\'\i}a-Fern{\'a}ndez, {\'A}ngel F.},
  journal = {IEEE Transactions on Automatic Control},
  title   = {Temporal Parallelization of Dynamic Programming and Linear Quadratic Control},
  volume = {68},
  issue = {2},
  pages = {851--866},
  year    = {2023},
}

@inproceedings{sarkka2023temporal,
  title={On The Temporal Parallelisation of The {V}iterbi Algorithm},
  author={S{\"a}rkk{\"a}, Simo and Garc{\'\i}a-Fern{\'a}ndez, {\'A}ngel F.},
  booktitle={2023 31st European Signal Processing Conference (EUSIPCO)},
  pages={2018--2022},
  year={2023},
}

@book{Sarkka+Svensson:2023,
       title = {Bayesian Filtering and Smoothing},
      author = {S{\"a}rkk{\"a}, Simo and Svensson, Lennart },
      edition = {2nd},
        year = {2023},
   publisher = {Cambridge University Press},
}

@article{sarkka2024temporal,
  title={Temporal Parallelisation of the {HJB} Equation and Continuous-Time Linear Quadratic Control},
  author={S{\"a}rkk{\"a}, Simo and Garc{\'\i}a-Fern{\'a}ndez, {\'A}ngel F.},
  journal={IEEE Transactions on Automatic Control},
  volume={70},
  number={6},
  pages={3755--3770},
  year={2024},
  publisher={IEEE}
}

@misc{sarkka2025perf,
  title={On The Performance of Prefix-Sum Parallel {K}alman Filters and Smoothers on {GPU}s},
  author={S{\"a}rkk{\"a}, Simo and Garc{\'\i}a-Fern{\'a}ndez, {\'A}ngel F.},
  note={arXiv:2511.10363},
  year={2025},
}

@inproceedings{todorov2008general,
  title={General duality between optimal control and estimation},
  author={Todorov, Emanuel},
  booktitle={2008 47th IEEE conference on decision and control},
  pages={4286--4292},
  year={2008},
}

@Article{Viterbi:1967,
  author = {Viterbi, Andrew~J. },
  title  = {Error bounds for convolutional codes 
            and an asymptotically optimum decoding
            algorithm},
  journal = {IEEE Transactions on Information Theory},
  volume = {13},
  number = {2},
    year = {1967},
}

@book{Van-Trees:1968,
      author = {{Van~Trees}, Harry. L.},
       title = {Detection, Estimation, and Modulation Theory, Part~I: Detection, Estimation, and Linear Modulation Theory},
   publisher = {John Wiley \& Sons},
     address = {New York, NY},
        year = {1968}
}

@book{Van-Trees:1971,
      author = {{Van~Trees}, Harry. L. },
       title = {Detection, Estimation, and Modulation Theory, Part~II: Nonlinear Modulation Theory},
   publisher = {John Wiley \& Sons},
     address = {New York, NY},
        year = {1971}
}

@book{Wymeersch2007Iterative,
  author    = {Wymeersch, Henk},
  title     = {Iterative Receiver Design},
  publisher = {Cambridge University Press},
  year      = {2007},
}

@inproceedings{yaghoobi2021parallel,
  title={Parallel iterated extended and sigma-point Kalman smoothers},
  author={Yaghoobi, Fatemeh and Corenflos, Adrien and Hassan, Sakira and S{\"a}rkk{\"a}, Simo},
  booktitle={ICASSP 2021-2021 IEEE International Conference on Acoustics, Speech and Signal Processing (ICASSP)},
  pages={5350--5354},
  year={2021},
}

@article{zeitouni1987maximum,
  title={A maximum a posteriori estimator for trajectories of diffusion processes},
  author={Zeitouni, Ofer and Dembo, Amir},
  journal={Stochastics: An International Journal of Probability and Stochastic Processes},
  volume={20},
  pages={221--246},
  year={1987},
  publisher={Taylor \& Francis}
}
